# Universal conductivity in the boson Hubbard model in a magnetic field

Min-Chul Cha

*Department of Physics and Astronomy and Center for Computational Sciences, University of Kentucky, Lexington, Kentucky 40506*

S. M. Girvin

*Department of Physics, Indiana University, Bloomington, Indiana 47405*

## Abstract

The universal conductivity at the zero-temperature superconductor-insulator transition of the two-dimensional boson Hubbard model is studied for cases both with and without magnetic field by Monte Carlo simulations of the (2+1)-dimensional classical $XY$-model with disorder represented by random bonds correlated along the imaginary time dimension. The effect of magnetic field is characterized by the frustration $f$. From the scaling behavior of the stiffness, we determine the quantum dynamical exponent $z$, the correlation length exponent $\nu$, and the universal conductivity $\sigma^*$. For the disorder-free model with $f = 1/2$, we obtain $z \approx 1$, $1/\nu \approx 1.5$, and $\sigma^*/\sigma_Q = 0.52 \pm 0.03$ where $\sigma_Q$ is the quantum conductance. We also study the case with $f = 1/3$, in which we find $\sigma^*/\sigma_Q = 0.83 \pm 0.06$. The value of $\sigma^*$ is consistent with a theoretical estimate based on the Gaussian model. For the model with random interactions, we find $z = 1.07 \pm 0.03$, $\nu \approx 1$, and $\sigma^*/\sigma_Q = 0.27 \pm 0.04$ for the case $f = 0$, and $z = 1.14 \pm 0.03$, $\nu \approx 1$, and $\sigma^*/\sigma_Q = 0.49 \pm 0.04$ for the case $f = 1/2$.

PACS numbers: 74.40,74.76

## I. INTRODUCTION

Through recent experimental studies in various two dimensional systems [1–9], the presence of a zero-temperature phase transition between the superconducting phase and the insulating phase has been convincingly suggested [10]. The transition has been tuned by changing parameters, such as the thickness of homogeneous films [1–3], the strength of charging energy in Josephson junction arrays [4], the strength of disorder [5], or the magnitude of the external magnetic field applied perpendicular to the two dimensional systems [6–9]. We view this transition as a boson localization problem, in which electron pairs are treated as point bosons small on the scale of the diverging phase correlation length near the transition. This picture is certainly suitable in granular superconductors and Josephson junction arrays where the electron-pair wavefunctions are well defined in each grain and the superconductivity is obtained by the establishment of a long-range phase correlation among them. Recently it has been proposed [11,12] that even in homogeneous films, the zero-temperature superconductor-insulator transition will be controlled by phase fluctuations of the electron-pair wavefunction. The picture is based on the assumption that below a mean-field temperature, electrons form pairs but the true superconducting temperature is lowered by the enhanced phase fluctuations in low dimensional systems. The size of the electron-pairs is finite at the transition. Hebard and Paalanen [7,13] reported some evidence that in homogeneous films, such as composite $InO_x$ films, the magnitude of the superconducting order parameter is finite at the transition, supporting the idea that the relevant fluctuations at the transition are the long wavelength twists in the phase of the pair wavefunction. We cannot rule out the possibility, however, that in these systems the transition might possibly be induced by the vanishing of the magnitude of the order parameter [14].

Based on the assumption that the transition is a continuous phase transition and that the phase fluctuations of the superconducting order parameter give the primary contribution to the singular part of the free energy, scaling theories have predicted some universal properties of the transition such as critical exponents [15,16] and dimensionless combinations of critical amplitudes [17]. In particular, the conductivity at the zero-temperature superconductor-insulator transition has been predicted [18] to be a universal number, which is analogous to the universal jump of the superfluid density at Kosterlitz-Thouless transition. Experimentally measured values of resistivity (or resistance per square) at very low temperature in the vicinity of the superconductor-insulator transition are close to the quantum resistance [1,4–6], $R_Q \equiv h/(2e)^2$, in various systems, even though no consensus concerning whether the temperature is low enough to reach the critical regime and whether the measured values are truly universal (*i.e.* independent of material parameters) has emerged yet. Theoretically a model of interacting bosons moving in the presence of disorder has been proposed to capture the appropriate universality class [15,18]. Considerable theoretical effort has been devoted to the study of the phase transition in this model by Monte Carlo simulations [19,20] and by renormalization group calculations [21,22]. Also the universal conductivity was calculated in the various versions of the model through different techniques [12,23–27]. Even though those calculations support the existence of a *finite* critical conductivity at the transition, the calculated values are either inconsistent among different studies or a factor of 2–3 away from experimental values. So far, theoretical studies mostly concentrated on the case without magnetic field. Here we want to study the phase transition and the universal conductivity in an interacting boson model in the presence of an external magnetic field. We also consider the effect of disorder given in the form of a short range repulsive random interaction.

This zero-temperature phase transition can be understood by adopting the charge-vortex duality picture [11]. Near the transition, vortices and anti-vortices are induced by quantum fluctuations. In the superconducting state, delocalized bosons condense whereas vortices are localized, yielding no dissipation. On the other hand, in the insulating phase, bosons are localized and vortices condense. According to this scenario, at the superconductor-insulator transition both bosons and vortices are expected to be mobile [12], yielding a finite conductivity because the motion of charged bosons carries current while that of vortices produces voltage. The external magnetic field changes the density of vortices and breaks time reversal symmetry of the vortex distribution thereby modifying the universality class.

In this work, we study the transition in a "phase-only" model. Here we argue that fluctuations of the magnitude of the superconducting order parameter are irrelevant near the transition because they cost more energy than the fluctuations of the phase. This picture implies that there are coarse-grained boson islands in which lots of bosons condense to have well-defined phase angles locally. These islands are connected through inter-island hopping. In this picture the phase angle is the only dynamical variable and the number operator and the phase operator are conjugate variables. The transition is determined by the competition of the density fluctuations and the phase angle fluctuations. Strong disorder and interactions in general suppress the density fluctuations locally and cause more fluctuations in phase angle. The external magnetic field causes a frustration in the phase couplings in this model.

More explicitly, we consider a lattice boson Hubbard model with random on-site interaction:

$$\mathcal{H} = \sum_i U_i(n_i - n_0 - \mu)^2 - \sum_{\langle ij \rangle}(t_{ij} b_i^\dagger b_j + t_{ji} b_j^\dagger b_i) \quad , \tag{1}$$

where $b_i$ ($b_i^\dagger$) is the boson annihilation (creation) operator at site $i$, $n_i$ is the boson number operator at site $i$, $U_i$ is the interaction between bosons at site $i$, $t_{ij}$ is the nearest neighbor hopping matrix element, and $n_0 + \mu$ ($n_0$ is an integer and $-1/2 < \mu < 1/2$) is the parameter which determines the average background boson density. In the presence of an external magnetic field, we have

$$t_{ij} = t_0 e^{iA_{ij}} \tag{2}$$

$$A_{ij} = \frac{2\pi}{\Phi_0} \int_{x_i}^{x_j} \vec{A}(\vec{x}) \cdot d\vec{x} \tag{3}$$

where $\Phi_0$ is the flux quantum and $\vec{A}(\vec{x})$ represents the vector potential. When there is a well defined phase angle $\theta_i$ at each site $i$,

$$n_i = \frac{1}{i} \frac{\partial}{\partial \theta_i} \tag{4}$$

and

$$b_i \approx \sqrt{n_0} e^{i\theta_i} \tag{5}$$

if $n_0 >> 1$. The model then reduces to the quantum rotor model

$$\mathcal{H} = \sum_i U_i (n_i - \mu)^2 - \sum_{\langle ij \rangle} J \cos(\theta_i - \theta_j - A_{ij}) \tag{6}$$

where $J = \sqrt{n_0} t_0$. Here we have shifted $n_i - n_0 \rightarrow n_i$. When $\mu = 0$, we have the corresponding (2+1)-dimensional classical action, $S[\theta]$, defined by

$$Z = \mathrm{Tr}_{\{\theta\}} e^{-\beta \mathcal{H}} = \mathrm{Tr}_{\{\theta\}} e^{-S[\theta]} \quad , \tag{7}$$

which is obtained by the path integral transformation [28]:

$$S[\theta] = -\sum_r [K_\tau(i) \cos(\Delta_\tau \theta_r) + K \cos(\Delta_x \theta_r) + K \cos(\Delta_y \theta_r)] \tag{8}$$

where $i$ and $\tau$ denote position on the spatial plane and the (imaginary) temporal axis respectively, $r \in (i, \tau)$ is a point in the (2+1)-dimensional lattice, and $\Delta_\tau \theta_r = \theta_{r+\hat{\tau}} - \theta_r$, etc. $K_\tau(i)$ is the coupling parameter along the (imaginary) temporal axis, depending only on the spatial position. Here we restrict the model to the particle-hole symmetric case $\mu = 0$, since otherwise the action is complex. The randomness of $U_i$ in Eq. (6) is transformed into the randomness of $K_\tau$. The frustration $f$, of the model is defined by the summation of the phase change around each plaquette so that

$$2\pi f = A_{ij} + A_{jk} + A_{kl} + A_{li} \quad . \tag{9}$$

This model is equivalent to a model high-$T_c$ superconductor with columnar defects along the magnetic field recently studied by Lee, Stroud, and Girvin [29].

In Section II, we describe the properties of the model, including scaling hypotheses necessary to analyze the numerical data. We then study the pure model, in which $K_\tau(i) = K$, with frustration f=1/2 and f=1/3 in Section III. A random-$U$ model where $K_\tau$ are random numbers bounded by $0 < K_\tau/K < 2$, with frustration f=0 and f=1/2 is also studied. Section IV is reserved for the conclusions.

## II. PROCEDURE

The phase transition of the model defined in Eq. (8), is studied by Monte Carlo simulations. We take simple cubic (2+1)-dimensional lattices whose sizes are systematically changed to fit data using finite-size scaling. Periodic boundary conditions are imposed and the ordinary Landau gauge is used for the phase change due to the magnetic field. We change the coupling strength parameter $K$ to tune the transition of the model.

The stiffness to a twist at the boundaries is measured to find the transition point and the critical exponents. The stiffness along a spatial direction is proportional to the superfluid density, and has a finite value in the superfluid phase but vanishes in the insulating phase. In a finite size system, however, it vanishes smoothly and there is a residual superfluid density at the transition point, whose magnitude depends on the size of the system. From the size dependency, therefore, we can extrapolate to find the asymptotic scaling behavior by changing the system size while keep the same geometry of the system. Since the model is anisotropic in the presence of the correlated disorder along the imaginary time dimension or the frustration in the spatial planes, the diverging behaviors of correlation lengths near critical point are not isotropic in general, but they are related through the dynamical scaling hypothesis

$$\xi_\tau \sim \xi^z \quad , \tag{10}$$

where $z$ is the quantum dynamical exponent. This implies that in order to simulate the model to find the asymptotic scaling in a finite size system in which correlation lengths are limited by the size of the system, we need to take lattices with

$$L_\tau = A_s L^z \quad , \tag{11}$$

where $A_s$ is the aspect ratio, and $L$ and $L_\tau$ are the sizes in the spatial directions and in the imaginary time direction, respectively. In this work, we use lattices with $L_x = L_y = L$ and $L_\tau$ which is determined by Eq. (11).

In order to determine the different sizes of lattice with the same aspect ratio, therefore, information about the value of $z$ is prerequisite. For the pure boson Hubbard model with the particle-hole symmetry, it is known that $z = 1$ [15]. Quenched disorder or long-range interactions, which alter the diverging behavior of correlation lengths, will change $z$ [15,18]. For the model we are considering, we expect that impurities hinder the correlation along the spatial directions so that the correlation length along the spatial directions will diverge more slowly than that along the imaginary time direction. This certainly implies that $z > 1$. The microscopic particle-hole symmetry property of the model ($\partial / i\partial\theta_j \leftrightarrow -\partial / i\partial\theta_j$ in Eq. (6)) implies that the occupation of the lowest delocalized eigenstate (*i.e.* the mobility edge state) does not change the total number of particles at the brink of the superconductor-insulator transition. Since the compressibility has a form

$$\kappa \sim \xi^{z-d} \quad , \tag{12}$$

this condition implies that $z < d$ where $d$ is the spatial dimension. Thus the model is different from that with random on-site chemical potential with a short-range interaction, where it is expected the compressibility of the system is finite even at the transition and $z = d$. For the model we consider, therefore, we expect

$$1 < z < 2 \quad . \tag{13}$$

Due to the lack of the information about an exact value of $z$ for the model we are studying, we have tried several numbers for $z$. We also find $z$ by directly measuring the correlation functions. The correlation functions along one spatial direction, say $x$, and along the imaginary time direction have asymptotic behavior [10,32]

$$\begin{aligned} G_x(r) &\sim r^{-y_x} \\ G_\tau(r) &\sim r^{-y_\tau} \end{aligned} \tag{14}$$

at the transition point, respectively, where $y_x = d - 2 + z + \eta$ and $y_\tau = (d - 2 + z + \eta)/z$. The value of $z$ is obtained from $z = y_x / y_\tau$. However in finite size systems with periodic boundary conditions, instead of Eq. (14) we use the form [32]

$$G_x(r) = C_x(r^{-y_x} + (L-r)^{-y_x})$$
$$G_\tau(r) = C_\tau(r^{-y_\tau} + (L_\tau - r)^{-y_\tau}) \quad (15)$$

where $C_x$ and $C_\tau$ are fitting constants. The results are discussed in the next section. Also one might expect the magnetic field perpendicular to the two-dimensional system to change the exponent $z$. We find that the external magnetic field has little effect on $z$ for the pure model and only weakly changes $z$ in the random-$U$ model.

The frequency-dependent stiffness is given by

$$\rho_{\mu\nu}(i\omega_n) = \frac{1}{L_x L_y L_\tau}[K_\nu \langle \epsilon_\nu \rangle \delta_{\mu\nu} - K_\mu K_\nu \langle J_\mu^*(i\omega_n) J_\nu(i\omega_n)\rangle + K_\mu K_\nu \langle J_\mu^*(i\omega_n)\rangle \langle J_\nu(i\omega_n)\rangle]_{\rm avg} \quad (16)$$

with

$$\epsilon_\nu = \sum_r \cos(\Delta_\nu \theta_r - A_r^\nu) \quad (17)$$

and

$$J_\nu(i\omega_n) = \sum_r \sin(\Delta_\nu \theta_r - A_r^\nu) e^{i\omega_n \tau} \quad (18)$$

where the frequency $\omega_n$ is the wavenumber in $\tau$-direction. In Eq. (16), $\langle \ldots \rangle$ represents the thermal average and $[\ldots]_{\rm avg}$ represents the ensemble average over the impurity configurations. In this work, the contribution of the last term in Eq. (16) drops out because $Lf$ is an integer in the pure model or because there is vortex-antivortex symmetry (*i.e.* $f = 0$ or $f = 1/2$) in the disordered model.

The finite-size scaling ansatz for $\rho_{xx}(0)$ is given by [12]

$$\rho_{xx}(0) = \frac{1}{L_\tau} \tilde{\rho}_{xx}(tL^{1/\nu}, L_\tau / L^z) \quad . \quad (19)$$

where $t = K - K^*$ ($K^*$ is the critical coupling), $\nu$ is the correlation length critical exponent ($\xi \sim t^{-\nu}$), and $\tilde{\rho}_{xx}$ is a universal scaling function. Near the transition,

$$L_\tau \rho_{xx}(0) = \tilde{\rho}_{xx}(0, L_\tau / L^z) + AtL^{1/\nu} + \ldots \quad (20)$$

where $\tilde{\rho}_{xx}(0, L_\tau/L^z)$ is an universal number, and $A$ is a non-universal constant. Similar behavior holds for $(L^2/L_\tau)\rho_{\tau\tau}(0)$. We use these properties to find the transition point, the quantum dynamical exponent, $z$, and the critical exponent, $\nu$, in the next section.

The conductivity is obtained through the Kubo formula [12]

$$\sigma_{\mu\nu} = 2\pi\sigma_Q \lim_{\omega_n \to 0} \frac{\rho_{\mu\nu}(i\omega_n)}{\omega_n} \tag{21}$$

where $\sigma_Q = (2e)^2/h$. It is a (scaling) dimensionless quantity and has a universal value at the transition. In practice, however, the finite size contribution, which gives the residual superfluid density, must be subtracted to find the universal behavior of the conductivity at the transition. Because of the frequency dependence and the finite-size contribution, we expect the conductivity has a scaling form [12]

$$\frac{\sigma(i\omega_n)}{\sigma_Q} = \frac{\sigma^*}{\sigma_Q} - c(\frac{\omega_n}{\omega_0} - \alpha\frac{\omega_0}{\omega_n L_\tau}) + \ldots \tag{22}$$

where $\omega_0$ is a given frequency such as the ultraviolet cutoff frequency. We set $\omega_0 = 2\pi/a$ in this work, where $a$ is the lattice constant. We take the parameter $\alpha$ which gives the smallest deviation among the interpolated scaling curves for the different size systems.

## III. RESULTS

### A. Pure model with $f = 1/2$ and $f = 1/3$

First, we consider the case of the pure boson model with frustration. We first have to check if the transition is first order or continuous in this case, because the scaling argument which supports the universal conductivity is based on the assumption of the continuous transition. A recent Monte Carlo calculation of the case $f = 1/6$ on a triangular lattice suggests that the transition is first order [30]. We use the Lee and Kosterlitz method [31] to check the nature of the transition at larger values of $f$. Fig. 1 is the probability distribution of energy, $p(e)$, for $f = 1/2$ and $f = 1/3$, where $p(e) = e^{-KE}/\mathrm{Tr}_{\{E\}}e^{-KE}$ and $e$ is the energy per volume. If the transition is first order, we expect a double-well type curve, in which an energy barrier between the two wells separates two phases. In finite systems, the energy barrier might be disguised by a finite size effect that will be reduced as the size of system increases so that the energy barrier increases as the size of system increases. We do not have any energy barrier increasing as the size of the system increases, and the result supports a continuous phase transition for both $f = 1/2$ and $f = 1/3$.

Fig. 2 shows the scaling behavior of $L_\tau \rho_{xx}(0)$ and $(L^2/L_\tau)\rho_{\tau\tau}(0)$ for the pure (2+1)-dimensional $XY$-model with frustration $f = 1/2$ for systems of different sizes. We assume that $z = 1$, which is supported by the correlation function calculation (see Fig. 3). The curves cross around $K^* = 0.707 \pm 0.001$. The slopes of the curves give us information about the correlation length critical exponent $\nu$. The inset of Fig. 2 is the finite-size scaling of $L_\tau \rho_{xx}(0)$ with respect to the single variable $(K - K^*)L^{1/\nu}$. With $1/\nu = 1.5 \pm 0.3$, good scaling is obtained. This value of $\nu$ is very close to $\nu$ in the unfrustrated three-dimensional $XY$ model [33]. The scaling behavior of the conductivity is shown in Fig. 4. From these curves, we obtain $\sigma^*/\sigma_Q = 0.52 \pm 0.03$. The error range is mostly due to the uncertainty in the critical coupling, $K^*$. Compared to the unfrustrated case where $\sigma^*/\sigma_Q = 0.285 \pm 0.02$ [12], the universal conductivity for the case $f = 1/2$ is almost doubled. This result was predicted by Granato and Kosterlitz [34] based on a picture of coupled $XY$-models [35]. For the fully frustrated case ($f = 1/2$), there are two degenerate ground state configurations. Thus when we construct an effective Ginzburg-Landau action, we need two complex fields (or two species of bosons) fluctuating around each ground state configuration. If we neglect the couplings between fields, we have the Gaussian model. The universal conductivity is proportional to the number of boson species in the Gaussian model. Since the $1/N$-expansion calculation of the universal conductivity [12] supports the notion that the Gaussian model is good as a first approximation, the universal conductivity of the fully frustrated case will be estimated to be doubled compared to the unfrustrated case. Our calculation supports this conjecture.

We also study the case with $f = 1/3$, obtaining that $K^* = 0.841 \pm 0.03$ and $\sigma^*/\sigma_Q = 0.83 \pm 0.06$. Again $z$ is assumed to be 1, which is consistent with a calcula-tion of the correlation function, and we find $1/\nu \approx 1.5$. The universal conductivity in this case is approximately tripled compared to the unfrustrated case, which is consistent with the existence of three different ground state configurations.

### B. Random-$U$ model

For the (2+1)-dimensional $XY$-model with random correlated bonds in the imaginary time direction, we have tried several values for the quantum dynamical exponent, $z$. Since the lattice sizes along the imaginary time direction determined by Eq. (11) are in general non-integer numbers, we adjust the aspect ratio to make the lattice sizes as close to integers as possible. More than 1000 disorder configurations are taken near the transition. When $f = 0$, we have the best scaling behavior near $z \approx 1.07$. Fig. 5 shows that, in this case, the scaling curves of $L_\tau \rho_{xx}(0)$ and $(L^2/L_\tau)\rho_{\tau\tau}(0)$ for systems with the sizes of $10 \times 10 \times 9.070, 12 \times 12 \times 11.023, 14 \times 14 \times 13, \text{and} 16 \times 16 \times 14.997$ cross around a transition point $K^* = 0.4465 \pm 0.025$. Since the stiffness depends weakly on the aspect ratio, these curves were obtained by linear interpolation of the data from the systems which have sizes corresponding to the nearby integers. Here the aspect ratio is $A_s = 0.772$. For significantly different $z$, either we cannot obtain the crossing behavior at all or $L_\tau \rho_{xx}(0)$ and $(L^2/L_\tau)\rho_{\tau\tau}(0)$ cross at different points. The error bound for the number $z$, which gives good scaling, is roughly $z = 1.07 \pm 0.03$. A slightly larger number was obtained by Lee *et al.* [29] in a similar model at $f = 1/4$, however their error bound is too big to meaningfully compare the numbers. The value of $z$ is also consistent with the direct calculation of correlation functions as shown in Fig. 6, where we find $z \approx 1.06$. The value of $\nu = 1.0 \pm 0.3$ is obtained from the scaling curve shown in the inset of Fig. 5, where the error bound is the rough range for good scalings. In Fig. 5, we used $1/\nu = 1.0$ and $K^* = 0.4465$. This number for $\nu$ is consistent with the simple extension of the Chayes lower bound condition [36] for the model with correlated disorder, which tells us that $\nu \geq 2/d$ where $d = 2$ is the number of dimensions in which the disorder occurs. The error bound on $\nu$ obtained in this fitting, however, is too big to confirm the condition for this model. Fig. 7 shows the curve of the resistivity for different couplings. Here we plot the resistivity instead of the conductivity. In a disordered boson model, Sørensen *et al.* [23] could fit their resistivity data in a Drude form. Even though our data expressed in the form of resistivity are not fit in a Drude form, better scalings are obtained from them than those obtained from the corresponding conductivity data. The curves are obtained from system sizes $10 \times 10 \times 9, 12 \times 12 \times 11, 14 \times 14 \times 13$, and $16 \times 16 \times 15$. Since the errors caused by the uncertainty of the critical coupling are dominant compared with those caused by the uncertainty of $z$, we use systems with nearby integer sizes. From those curves, we determine the universal conductivity of the model: $\sigma^*/\sigma_Q = 0.27 \pm 0.04$.

With frustration $f = 1/2$ in the spatial planes of the model, we have $z \approx 1.14 \pm 0.03$, $K^* = 0.6875 \pm 0.025$, $\nu = 1.0 \pm 0.3$ (see Fig. 8), and $\sigma^*/\sigma_Q = 0.49 \pm 0.04$ (see Fig. 9). Again, the error bound on the values of $z$ and $\nu$ are the rough range for good fitting. The correlation function calculation gives $z \approx 1.17$. The universal conductivity in this case is also approximately doubled compared with the unfrustrated model. This means that the type of disorder we consider does not wash out the magnetic length. As for the critical coupling, even though the shape of the lattices used are slightly different in the simulations with and without randomness along the imaginary time axis, the critical coupling is consistently suppressed in the disordered case. This result is consistent with the recent simulation [29] of a model high-$T_c$ superconductor with columnar defects along the $c$-axis, where the $T_c$ is increased because of the vortex pinning effect due to the random disorder.

## IV. CONCLUSIONS

We have calculated the universal conductivity for a random-$U$ boson Hubbard model. For the fully frustrated ($f = 1/2$) pure system, we obtain $\sigma/\sigma_Q = 0.52 \pm 0.03$. The conductivity is almost doubled, as predicted by Granato and Kosterlitz [34], compared with the unfrustrated system. For $f = 1/3$, we find $\sigma/\sigma_Q = 0.83 \pm 0.06$ so that the conductivity is approximately tripled. The phase transition in the above two cases is continuous, and we find $z \approx 1$ and $1/\nu \approx 1.5$. For the disordered case, we study the model for $f = 0$ and $f = 1/2$. We obtain the quantum dynamical exponents $z = 1.07 \pm 0.03$ and $z = 1.14 \pm 0.03$, respectively. The critical exponent $1/\nu \approx 1$ is obtained, the value of which is consistent with Chayes lower bound condition [36]. The universal conductivity is $\sigma/\sigma_Q = 0.27 \pm 0.04$ and $\sigma/\sigma_Q = 0.49 \pm 0.04$ for $f = 0$ and $f = 1/2$, respectively. Disorder in the model has little effect on the value of the universal conductivity (at least at $f = 1/2$). Also it does not wash out the length scale set by the frustration, although the critical exponents are changed.

## ACKNOWLEDGMENTS

We appreciate many discussions with Dr. Mats Wallin. MCC is supported in part by NSF # DMR 92-02255 and the Center for Computational Sciences at University of Kentucky. SMG is supported by DOE # DE-FG02-90ER45427.

## FIGURES

FIG. 1. Probability distribution of energy in Monte Carlo simulation of the pure boson model for the case with $f = 1/2$ and $f = 1/3$ for different size systems indicated in the figure. The absence of an energy barrier separating two phases shows that the transition is continuous.

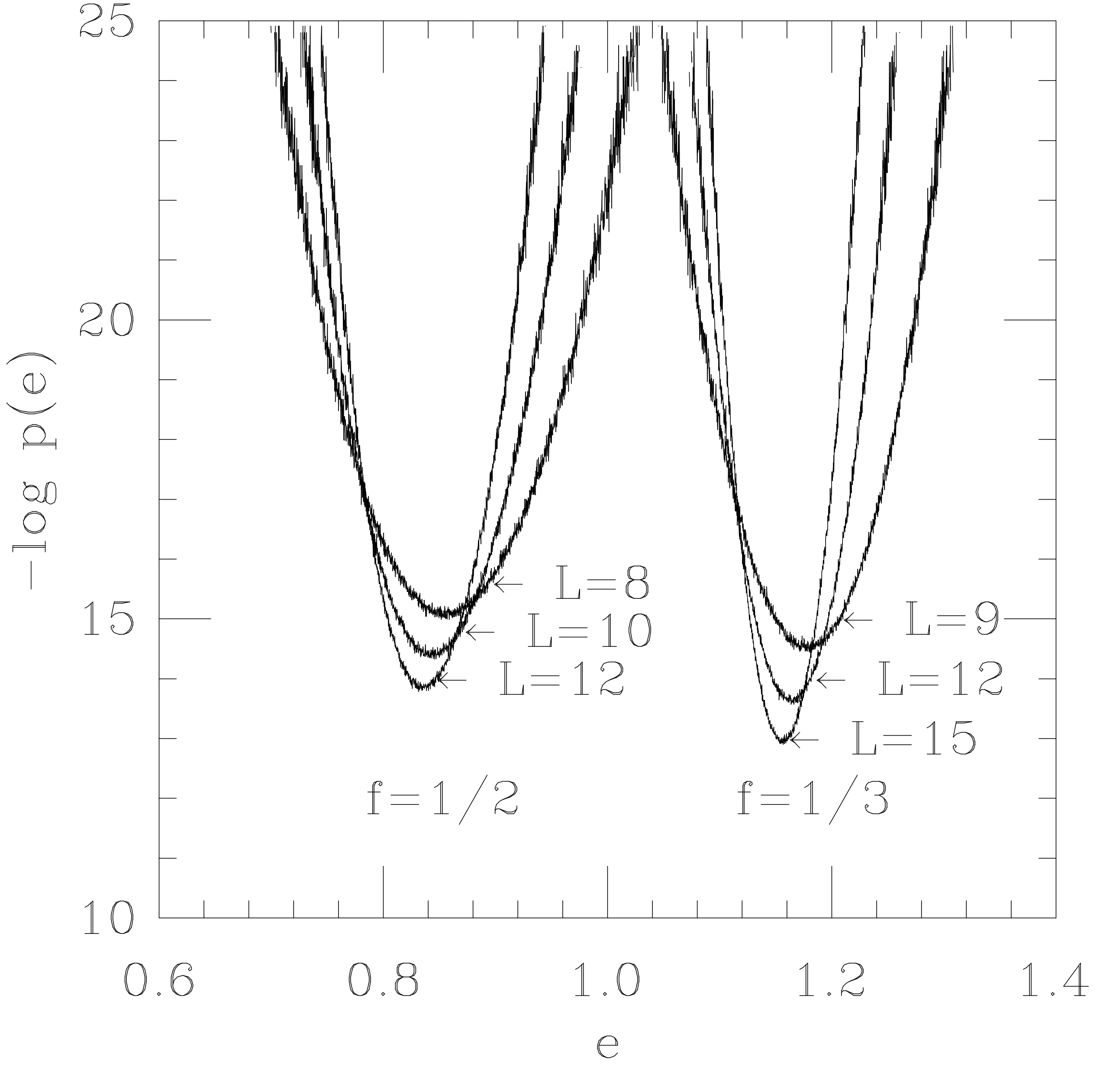

FIG. 2. Finite-size scaling behavior of $L_\tau \rho_{xx}(0)$ and $(L^2/L_\tau)\rho_{\tau\tau}(0)$ for the pure (2+1)-dimensional $XY$-model with $f = 1/2$. The transition point determined from these crossing curves is $K^* = 0.707 \pm 0.001$. Inset: The scaling behavior of $L_\tau \rho_{xx}(0)$ with respect to the single variable $(K - K^*)L^{1/\nu}$. Here we use $K^* = 0.7068$ and $1/\nu = 1.5$.

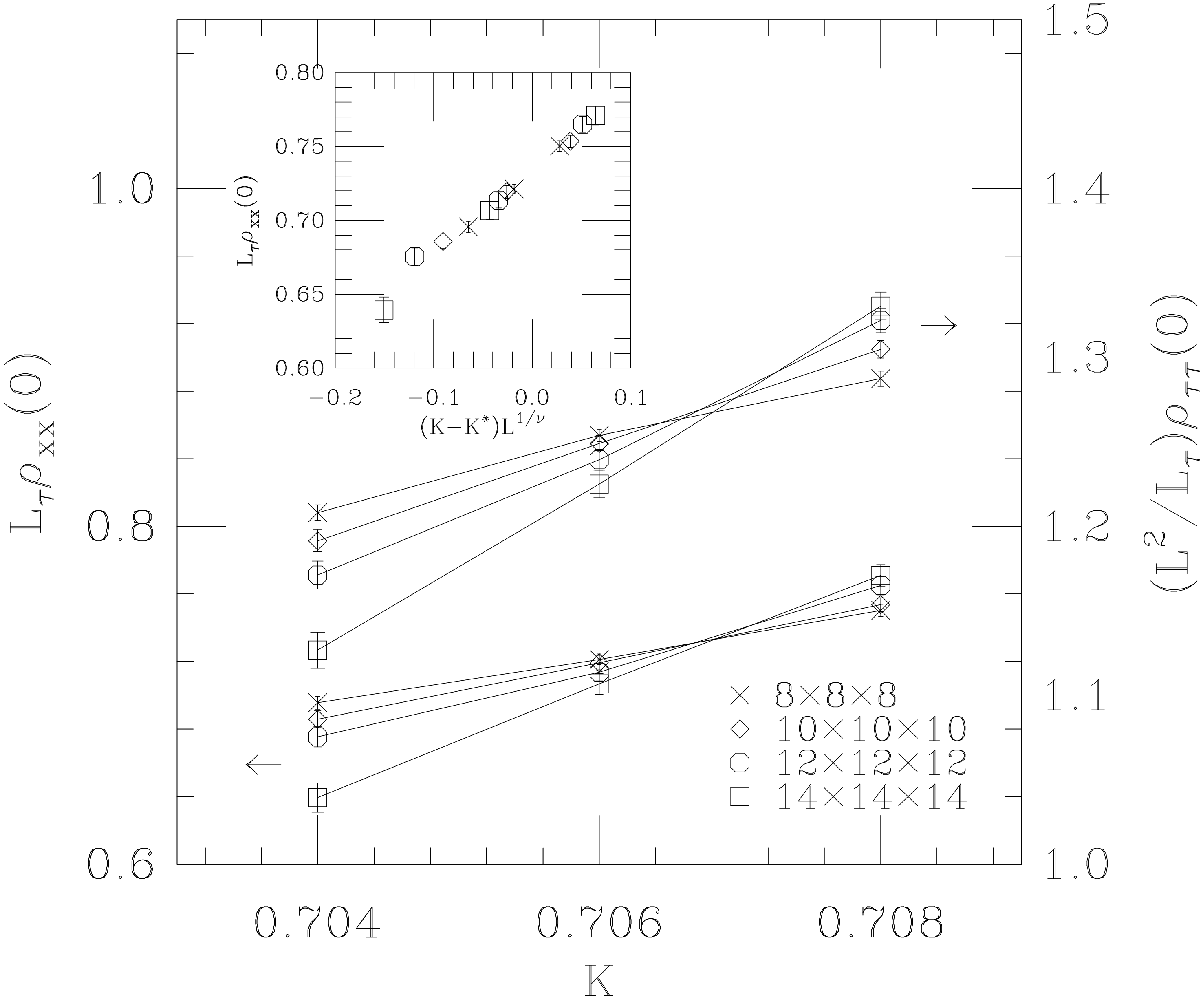

FIG. 3. Spin-spin correlation function of the pure (2+1)-dimensional $XY$-model along one spatial axis, say $x$, and along the imaginary time axis for frustration $f = 1/2$. ($\times$) and ($+$) represent actual data points and the solid lines represent the interpolated curves. Because of the frustration the spin-spin correlation function along a spatial axis oscillates slightly. The interpolated curve of $G_x(r)$ is the correlation function for spins separated by even numbers of lattice constants. The parameters used in the interpolation are $C_x = 0.4495, y_x = 1.071, C_\tau = 0.5341$, and $y_\tau = 1.075$. From the ratio of $y_x$ and $y_\tau$, the dynamical exponent is determined to be $z \approx 1$.

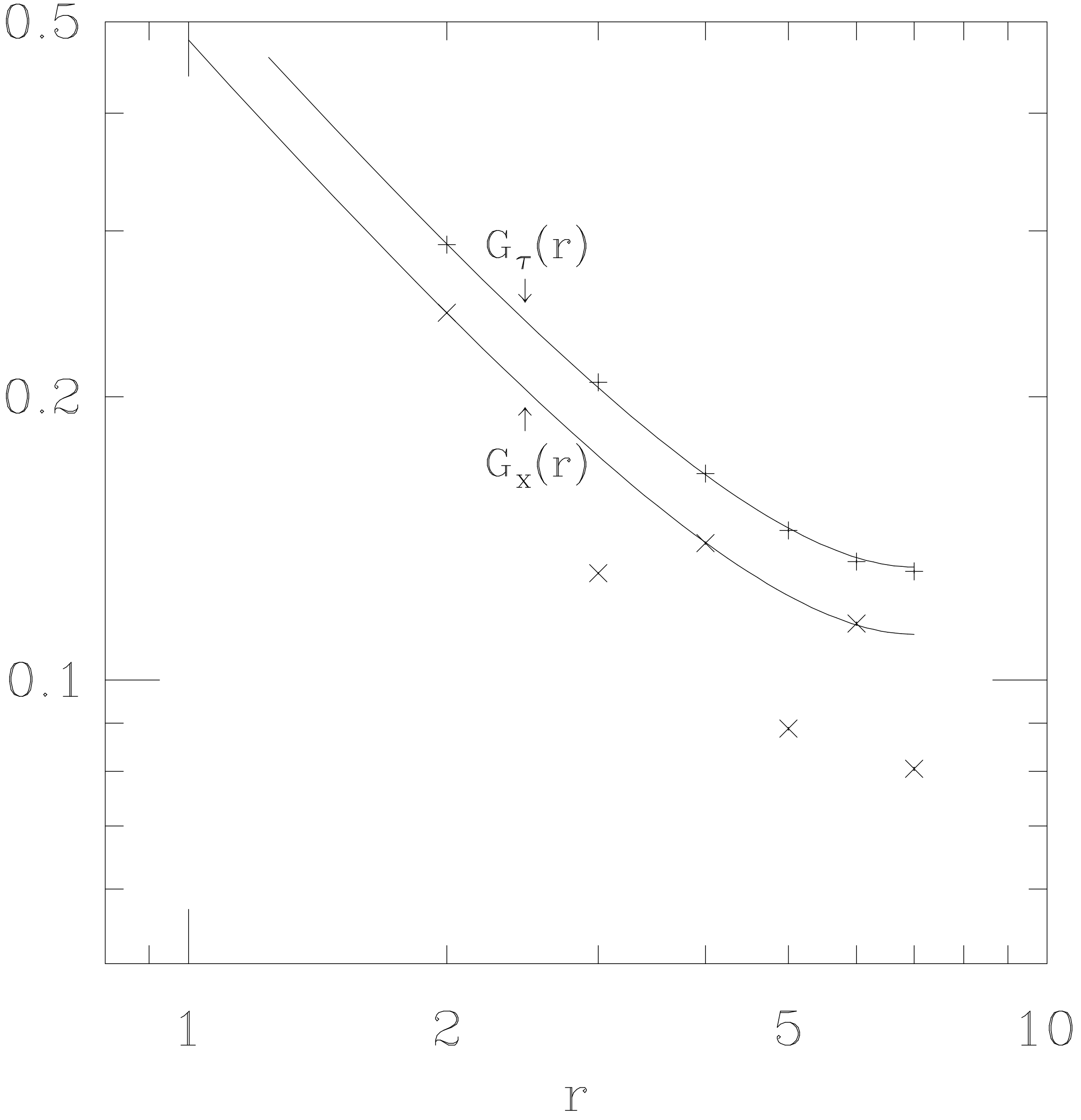

FIG. 4. Finite-size scaling curves of $\sigma(i\omega_n)/\sigma_Q$ for the pure (2+1)-dimensional $XY$-model. From the top, the sets of curves represent the interpolated curves for $K = 0.704, 0.706$, and $0.708$ respectively obtained for different size systems. Here we choose $\alpha = 0.68, 0.61$, and $0.48$, respectively. From these curves, the d.c. conductivity is determined by the value at $\omega_n/\omega_0 - \alpha\omega_0/(\omega_n L_\tau) = 0$. The universal conductivity obtained at the transition point is $\sigma^*/\sigma_Q = 0.52 \pm 0.03$ in this case.

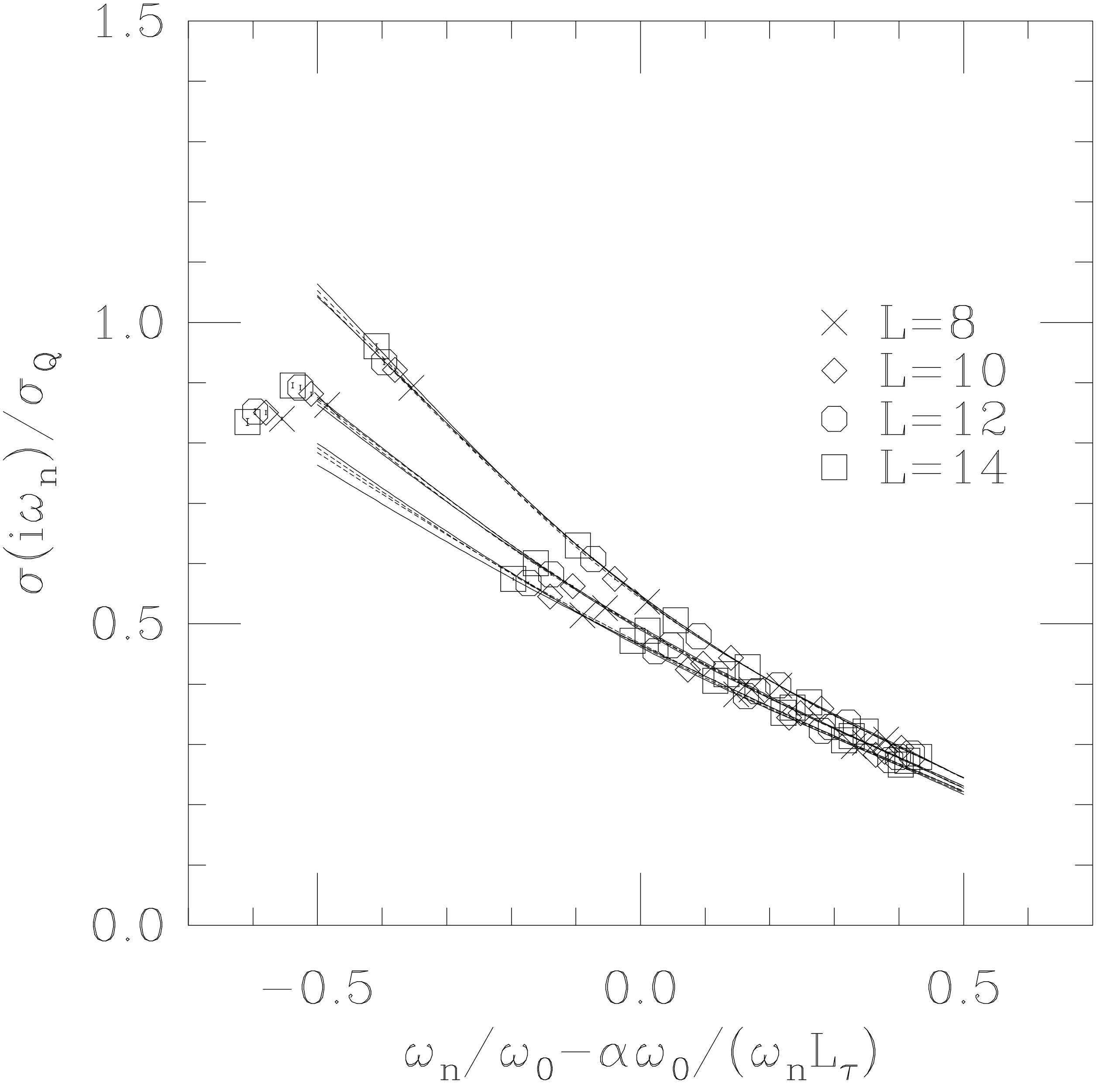

FIG. 5. Finite-size scaling behavior of $L_\tau \rho_{xx}(0)$ and $(L^2/L_\tau)\rho_{\tau\tau}(0)$ for the (2+1)-dimensional $XY$-model with correlated random bonds along the imaginary time axis. The data points for the systems with non-integral size are obtained by the linear interpolation of the data obtained for systems which have nearby integer sizes. The transition point determined from these crossing curves is $K^* = 0.4465 \pm 0.025$. Inset: The scaling behavior of $L_\tau \rho_{xx}(0)$ with respect to the single variable $(K - K^*)L^{1/\nu}$. $K^* = 0.4465$ and $1/\nu = 1.0$ are used.

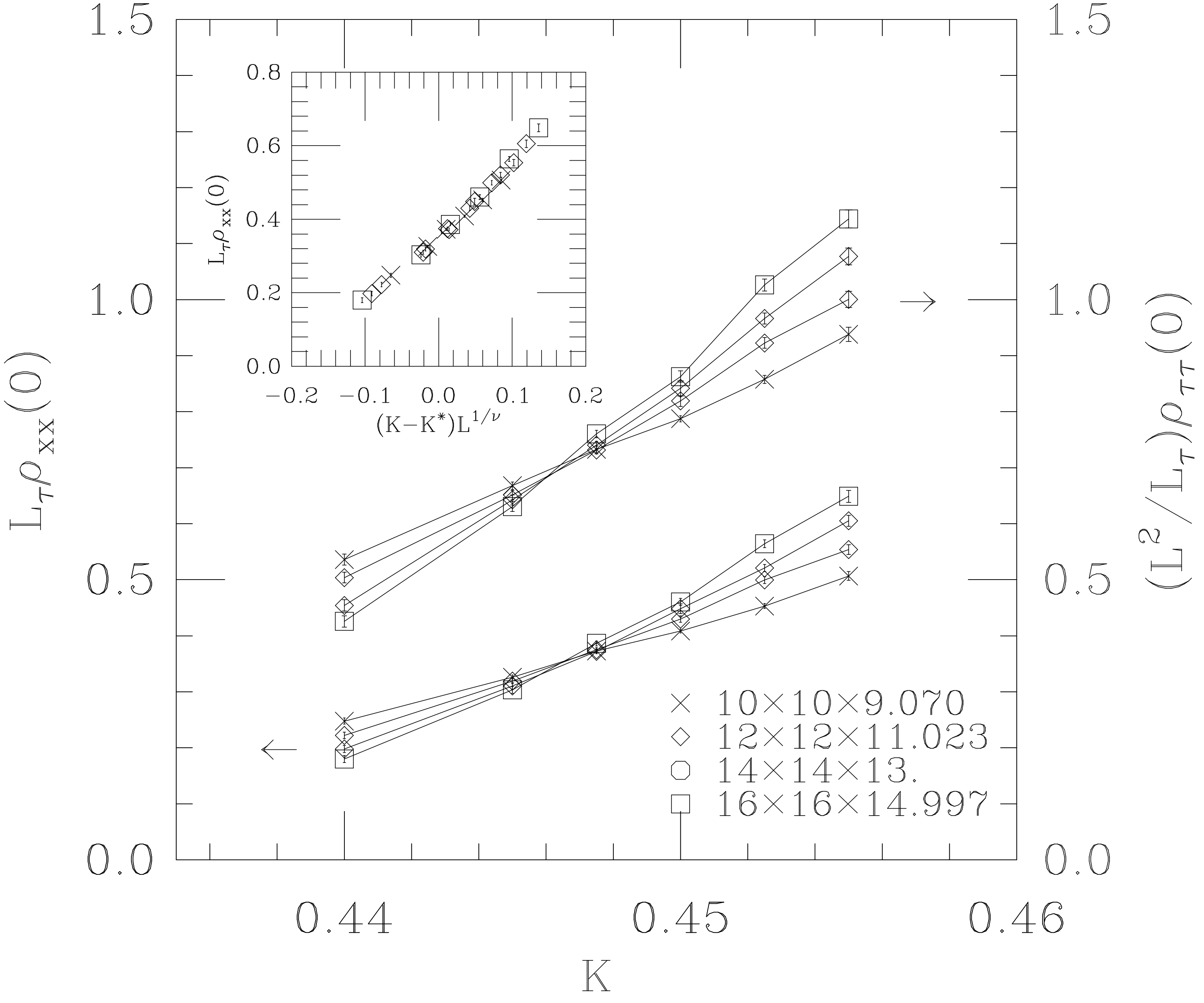

FIG. 6. Spin-spin correlation function of the (2+1)-dimensional $XY$-model along one spatial axis, say $x$, and along the imaginary time axis. ($\times$) and (+) represent actual data points and the solid lines represent the interpolated curves. The parameters used in the interpolation are $C_x = 0.2865, y_x = 1.054, C_\tau = 0.2975$, and $y_\tau = 0.994$. From the ratio of $y_x$ and $y_\tau$, the dynamical exponent is determined to be $z \approx 1.06$.

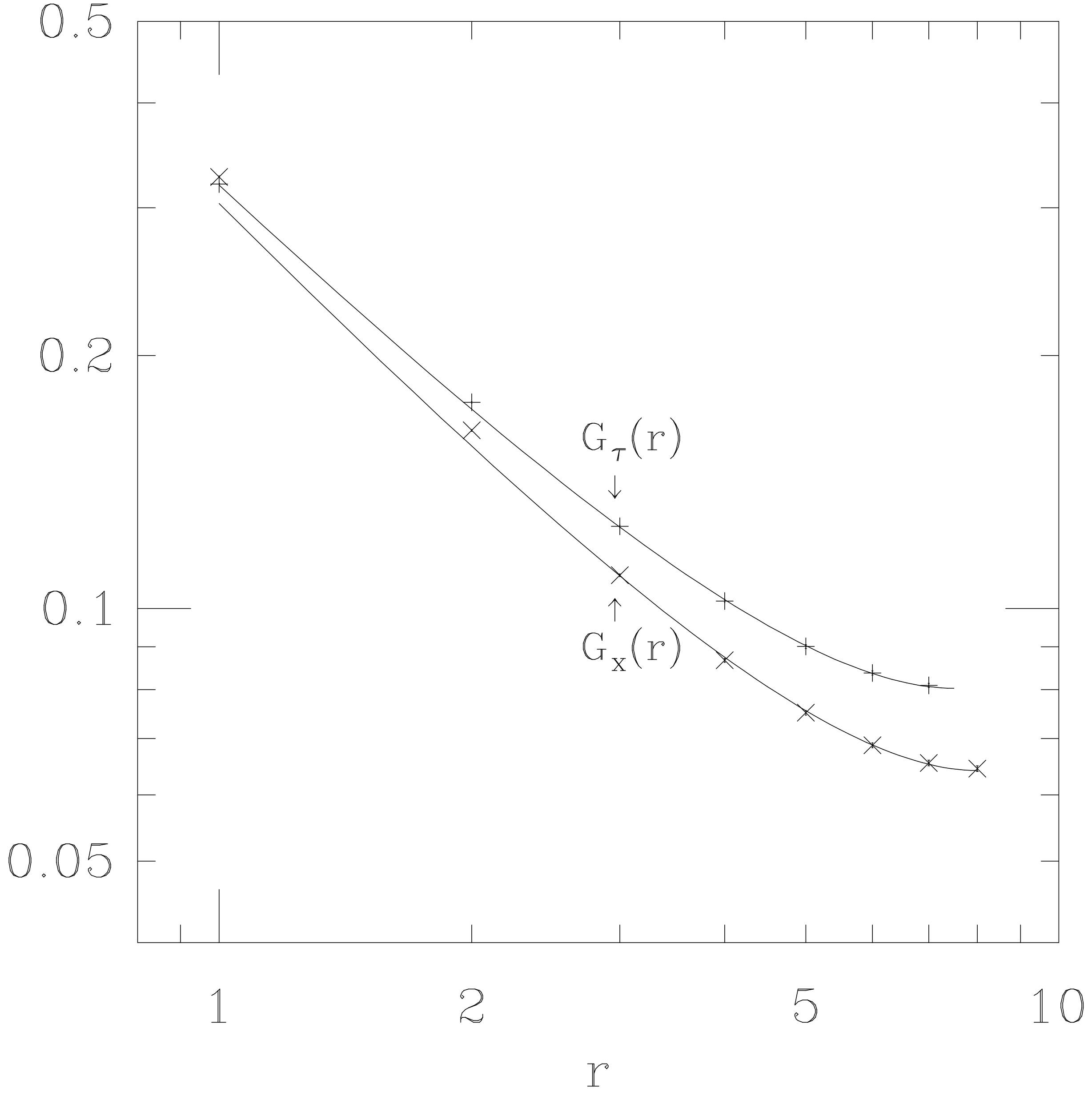

FIG. 7. Finite-size scaling curves for the resistivity, $R(i\omega_n)/R_Q$, for the (2+1)-dimensional $XY$-model with random bonds correlated along the imaginary time axis. From the top, the sets of curves represent the interpolated curves for $K = 0.440, 0.445, 0.4475, 0.450$, and $0.455$ respectively. We choose $\alpha = 0.88, 0.55, 0.39, 0.28$, and $0.16$, respectively. The universal conductivity determined from these curves is $\sigma^*/\sigma_Q = 0.27 \pm 0.04$.

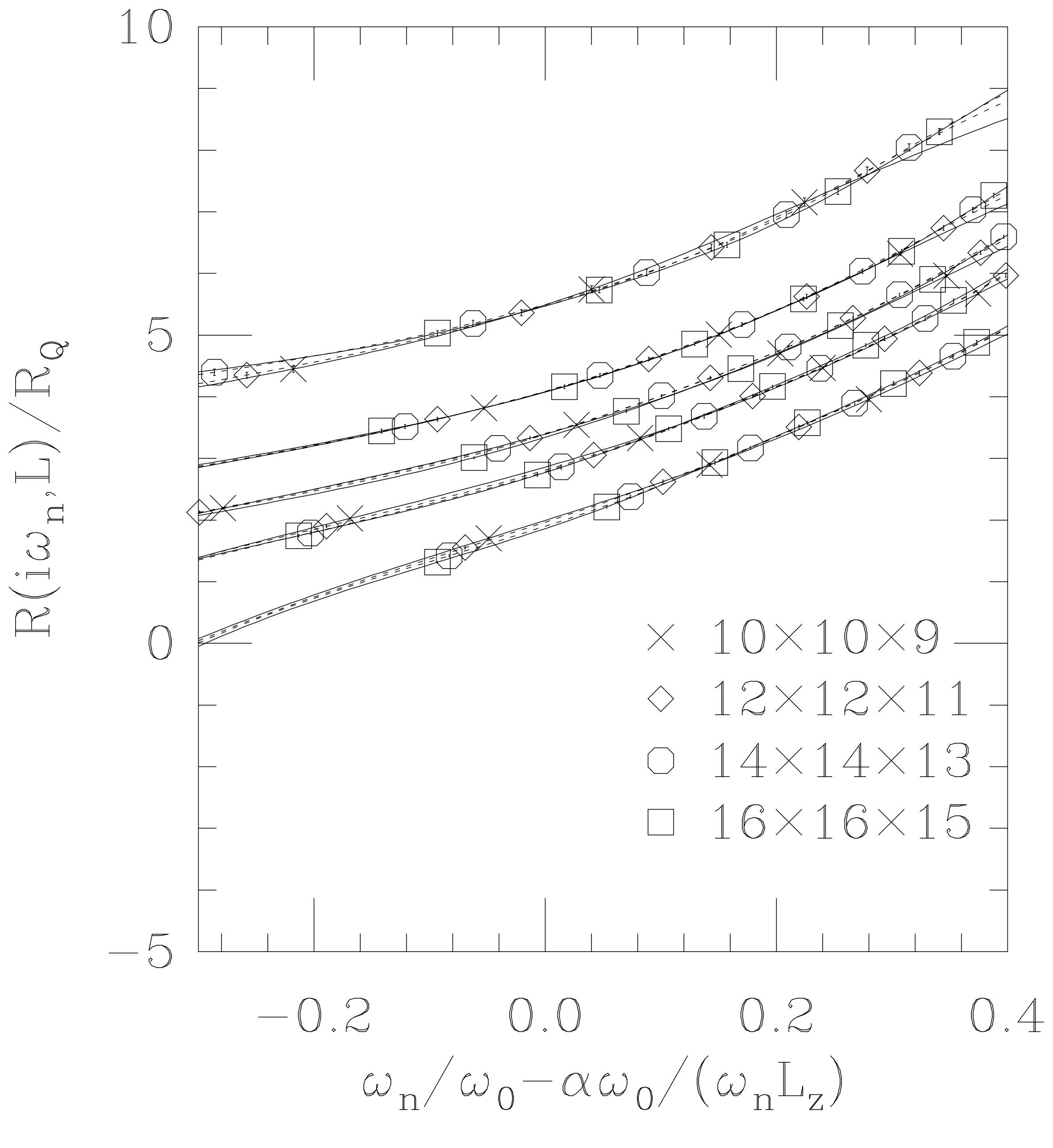

FIG. 8. Finite-size scaling behavior of $L_\tau \rho_{xx}(0)$ and $(L^2/L_\tau)\rho_{\tau\tau}(0)$ for the (2+1)-dimensional $XY$-model with correlated random bonds along the imaginary time axis with frustration $f = 1/2$ in the spatial plane. The transition point determined from these crossing curves is $K^* = 0.6875 \pm 0.025$. Inset: The scaling behavior of $L_\tau \rho_{xx}(0)$ with respect to the single variable $(K - K^*)L^{1/\nu}$. $K^* = 0.6875$ and $1/\nu = 1.0$ are used.

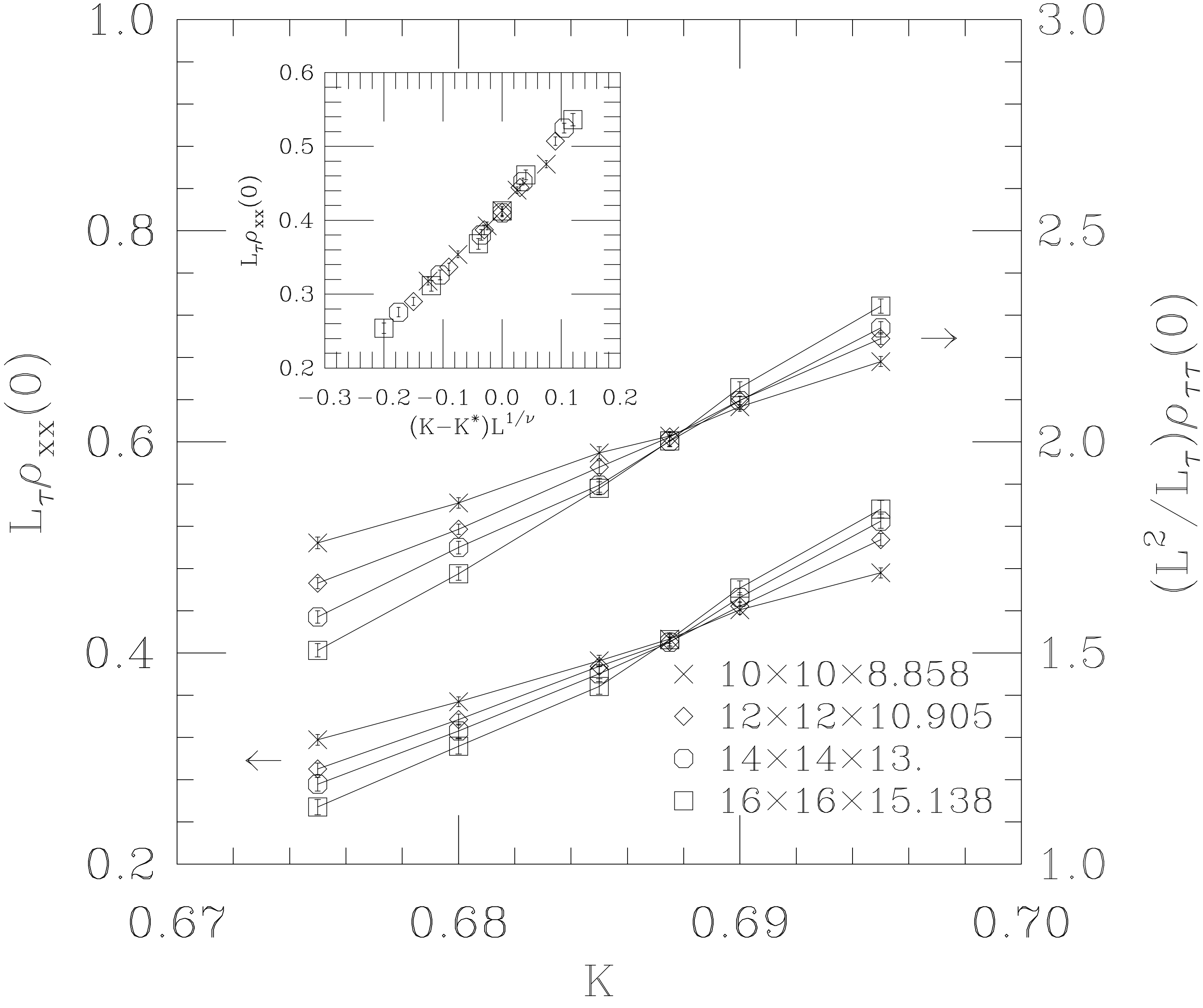

FIG. 9. Finite-size scaling curves of the resistivity, $R(i\omega_n)/R_Q$, for the (2+1)-dimensional $XY$-model with random bonds correlated along the imaginary time axis with $f = 1/2$. From the top, the sets of curves represent the interpolated curves for $K = 0.670, 0.680, 0.685, 0.6875, 0.690,$ and $0.695$. We choose $\alpha = 0.40, 0.35, 0.30, 0.27, 0.24,$ and $0.20$, respectively. The universal conductivity determined from these curves is $\sigma^*/\sigma_Q = 0.49 \pm 0.04$.

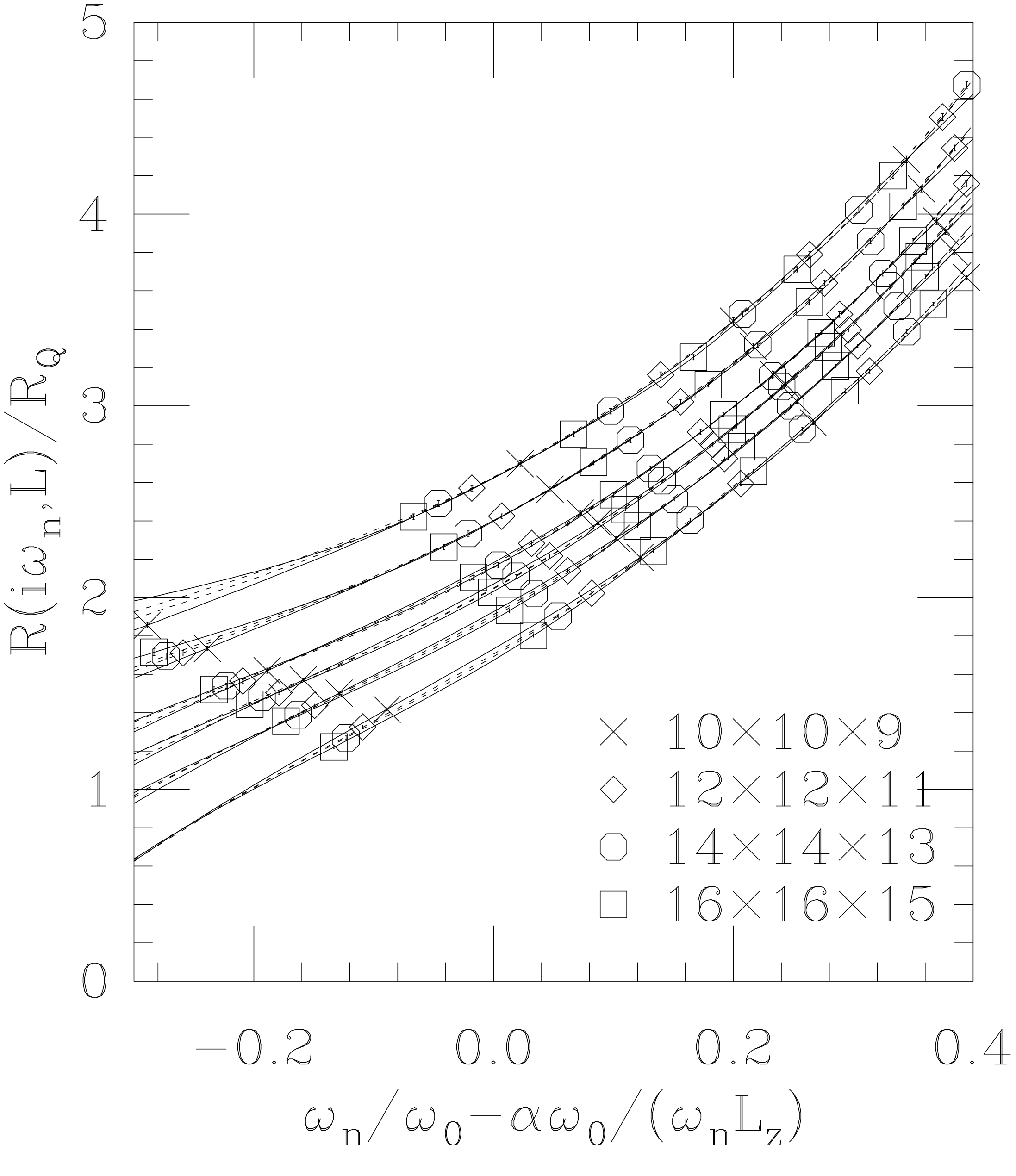